%% file: main.tex
\documentclass[journal]{IEEEtran}
\usepackage[T1]{fontenc}
\usepackage[latin9]{inputenc}
\usepackage{amsmath}
\usepackage{amssymb}
\usepackage{amsthm,bm}
\usepackage{url}
\usepackage[table]{xcolor}  
\usepackage{tabularx}
\usepackage{pifont}% http://ctan.org/pkg/pifont
\newcommand{\cmark}{\ding{51}}%
\theoremstyle{plain}

\theoremstyle{remark}

\makeatother

\usepackage{babel,verbatim,balance}
\usepackage{tcolorbox}
\usepackage{graphics, graphicx}
\usepackage{upgreek}
\usepackage{setspace,gensymb}

\input{acronyms}

\begin{document}
\bstctlcite{IEEEexample:BSTcontrol}
\title{Towards Distributed and Intelligent Integrated Sensing and Communications for 6G Networks}
\author{Emilio Calvanese Strinati, George C. Alexandropoulos, Navid Amani,  Maurizio Crozzoli, \\ Giyyarpuram Madhusudan, Sami Mekki, Francois Rivet, Vincenzo Sciancalepore, Philippe Sehier, \\ Maximilian Stark, and  Henk Wymeersch
\thanks{This work has been supported by the SNS JU project 6G-DISAC under the EU's Horizon Europe research and innovation Program under Grant Agreement No 101139130.}
}

\maketitle

\begin{abstract}
This paper introduces the \ac{DISAC} concept, a transformative approach for 6G wireless networks that extends the emerging concept of \ac{ISAC}. \ac{DISAC} addresses the limitations of the existing \ac{ISAC} models and, to overcome them, it introduces two novel foundational functionalities for both sensing and communications: a distributed architecture (enabling large-scale and energy-efficient tracking of connected users and objects, leveraging the fusion of heterogeneous sensors) and a semantic and goal-oriented framework (enabling the transition from classical data fusion to the composition of semantically selected information). 
\end{abstract}
%\begin{IEEEkeywords}
%6G, ISAC, Distributed, Semantic.
%\end{IEEEkeywords}

\acresetall 

\section{Introduction}
The advent of 6G brings a fundamental rethinking of the purpose of radio signals and new technologies, from communications to sensing \cite{ISAC_IEEE_Netw}. In such context, \ac{ISAC} emerges as a pivotal domain of research and development {in the forthcoming 6G wireless networks}. In fact, \ac{ISAC} promises to usher in a new era of connectivity, where communication is not limited to data transfer, but extends its reach into sensing, knowledge, intelligence, and reconfiguration, thereby connecting the physical and digital worlds. % \cite{uusitalo20216g}. 

Despite the great progress on its fundamentals, \ac{ISAC} has often remained at low \acp{TRL}, presenting challenges and barriers that need to be surmounted. In particular, the current vision for \ac{ISAC} generally misses three critical components: \textit{i}) it does not provide support for widely distributed deployments, to track both many connected \acp{UE} and many passive objects over extended time and space; \textit{ii}) %it focuses on the performance in terms of the \acp{KPI}, rather than the more holistic \acp{KVI};
it focuses solely on the performance in terms of the \acp{KPI}, while 6G must adopt a comprehensive perspective that incorporates both \acp{KPI} and \acp{KVI};
\textit{iii}) it focuses only on the 6G signal as a sensor, ignoring the fusion opportunities with external sensors, as well as more general semantic awareness. Unlocking the potential of \ac{ISAC} requires us to extend that vision, by addressing the latter three critical components. In this paper, we propose such an extended vision, termed \emph{\acf{DISAC}}. 

\begin{figure}
    \centering
    \includegraphics[width=\columnwidth]{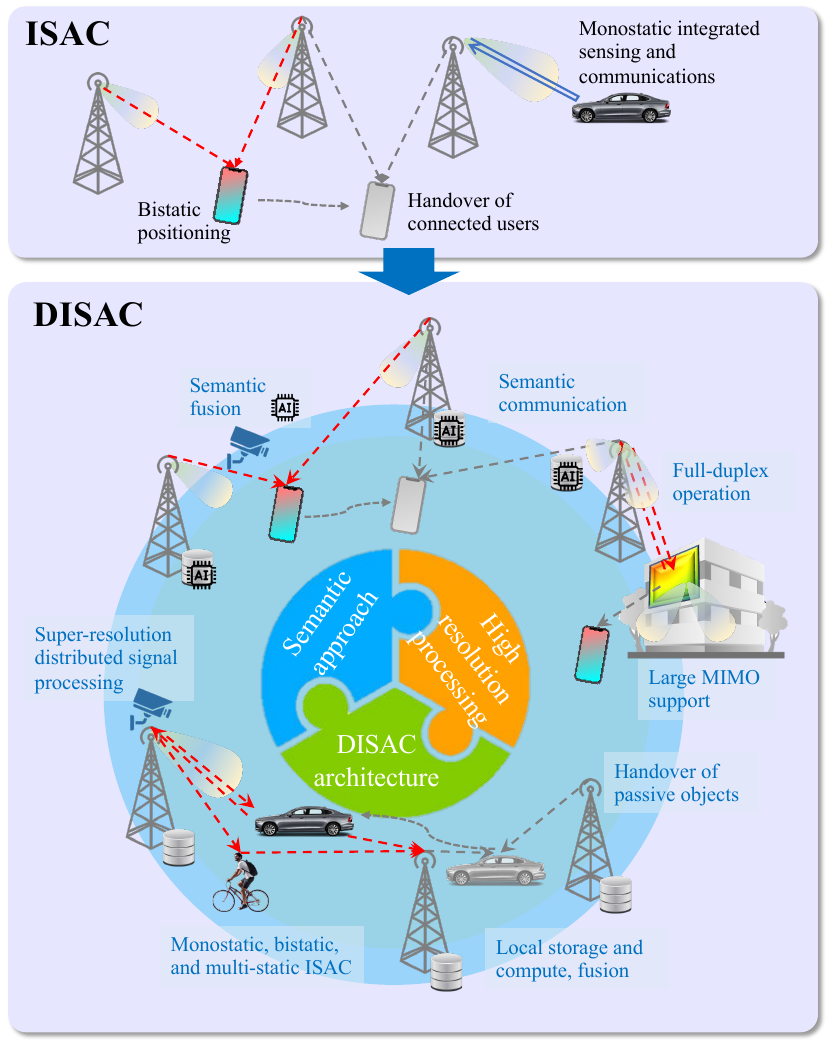}
    \caption{The vision evolution from \ac{ISAC} to \ac{DISAC}, the latter being implemented by a new architecture that supports an \ac{AI}-based semantic approach and super-resolution multi-sensor processing.}
    \label{fig:DISAC-section1}
\end{figure}

The \ac{DISAC} framework for 6G represents a transformative vision for wireless communications, 
combining the fusion of heterogeneous and distributed sensors with a highly adaptive and efficient semantic-native approach to enable energy-efficient, high-resolution tracking of connected \acp{UE} and objects.
\Ac{DISAC} is built on several interrelated cornerstones, visualized as puzzle pieces in Fig.~\ref{fig:DISAC-section1}. 
First is the \ac{DISAC} architecture, which serves as the foundation for both sensing and communications, while simultaneously offering support for intelligent operations and distributed functions. This distributed aspect, not only enables large-scale tracking of connected \acp{UE} and passive objects, but also revolutionizes the fundamental fabric of wireless networks. 
{The DISAC architecture 
supports distributed operations enabled/improved by \ac{AI},  
with careful balancing of centralized vs.~local data processing and fusion of information from different remote sensors, }
novel multi-antenna technologies, and an exposure framework for external sensors. 
{The second cornerstone is the semantic and goal-oriented framework, 
providing an intelligent and parsimonious framework encompassing sensing activation, waveform design, signal processing, dedicated resource allocation, robust protocols, selective information sharing (semantic information, extracted and processed given accumulated background knowledge)
as well as reasoning about multi-modal sensed information (i.e., generated by heterogeneous sensor types), to enable continuous and multi-modal learning and action-taking (semantic reasoning). This framework, supported by \ac{ML} and \ac{AI}, ensures exceptional sensing performance for a myriad of use cases while optimizing resource utilization.}
The third and final cornerstone is advanced high-resolution processing, taking advantage of the massively distributed observations while balancing computational and storage requirements. By exploiting multi-site and multi-band processing and by leveraging a combination of \ac{ML} and model-based signal processing, efficient methods that can support the goal-oriented/semantic framework while running over the \ac{DISAC} architecture must be developed and verified.

Deploying \ac{DISAC} on a global scale requires standardization efforts, especially concerning \ac{AI}/\ac{ML}-driven sensing within distributed heterogeneous architectures. 
It would also impact various layers of the 6G ecosystem, including the physical layer, the control and management planes, and the security of data and model exchanges. Furthermore, achieving harmonious orchestration of radio, transport, and processing resources becomes of primary importance, demanding efficient and dynamic solutions.
{In this paper, we detail our vision for the DISAC framework, relying on a combination of proven technologies from different fields, as well as promising, yet untested concepts, such as the semantic framework, and bringing together perspectives from the telecommunications industry, key verticals, and academia. }

\section{Use Cases and Standardization}

In this section, we explain the necessity for DISAC from two perspectives: use cases and standardization.

\subsection{The Pull of Use Cases}
\label{sec:use-cases}
{Over 30 potential use cases of interest for both industry and the public sector, enabled by integrated sensing and 5G/6G communication technology, are proposed in~\cite{3GPPTR22837}.
Those use cases and their  envisioned associated 6G services {\cite[Section 2.2]{CalvaneseGOWSC2021}} 
are analyzed and related to the three DISAC cornerstones  (the \textit{architecture}, 
the
\textit{semantic approach} 
and the \textit{high-resolution processing}). 
 From this analysis, summarized in Table I, it is evident that,}
in all use cases that cover extended environments, DISAC is necessary to support high-precision spatial-temporal processing and prediction beyond the view of a single sensor, to balance communication, computational, and storage loads among the network entities, and to provide actionable intelligence, based on reasoning and compounding knowledge from diverse perspectives. Moreover, use cases that can be enhanced by the fusion of multi-modal sensed information can benefit from a semantic type of information exchange and fusion.

\newcolumntype{P}[1]{>{\centering\arraybackslash}m{#1}}
\newcolumntype{Q}[1]{>{\arraybackslash}m{#1}}
\begin{table*}[]
    \centering
    \resizebox{1.8\columnwidth}{!} {
    {
    \begin{tabular}
{|Q{7cm}|P{0.2cm}|P{0.2cm}|P{0.2cm}|P{0.2cm}|P{2cm}|P{2cm}|P{2cm}|}
    \hline
    \rowcolor{blue!10}    
     \textbf{Use Case from~\cite{3GPPTR22837} (sections)}
     & \rotatebox{90}{\textbf{MMTCCxDI}~} & \rotatebox{90}{\textbf{GeMBB}} & \rotatebox{90}{\textbf{URLLCCC}} & \rotatebox{90}{\textbf{Semantic}}
     &\textbf{DISAC Architecture} (D) distributed processing; (W) wide perspective over space and time; and (S) support for non-3GPP sensors & \textbf{DISAC Semantic Approach} (R) resource-efficient operation (including compression of data); (A) \ac{AI}-based and semantic reasoning; and (M) multi-modal sensing & \textbf{DISAC High-Resolution Processing} (D) distributed sensing and (E) extreme requirements \\
    \hline
    \hline
    Intrusion detection in/around smart homes  (5.1, 5.6, 5.16) & \cmark &  &\cmark  & \cmark  & \cmark (S) & \cmark (R, A, M) &   \\
    \hline
    Highway, railway intrusion detection (5.2, 5.7) &  & \cmark &  &    & \cmark (D, W) &  \cmark (A) & \cmark (E)   \\
    \hline
    Rainfall monitoring (5.3) & \cmark & \cmark &  &  & \cmark (D), \cmark (W) & \cmark (A) & \cmark (D) \\
    \hline
    Transparent sensing ( 5.4) & \cmark &  & \cmark &   & \cmark (S)  & \cmark (R, A, M) &   \\
    \hline
    Sensing for flooding  (5.5) & \cmark & \cmark &  & \cmark &  \cmark (D, W) & \cmark (A) & \cmark (D)   \\
    \hline
    Automotive maneuvering and navigation  (5.8, 5.26, 5.30) & \cmark &  & \cmark & \cmark  & \cmark (D, W, S) & \cmark (R, A, M) & \cmark (D, E)   \\
    \hline
AGV detection and tracking in factories  (5.9) & \cmark  & \cmark & \cmark &   &   \cmark (D, W) &  &  \cmark (D)  \\
    \hline
    UAV trajectory tracing and intrusion detection   (5.10, 5.13, 5.22) &  & \cmark & \cmark & \cmark&  \cmark (D, W, S) & \cmark (R, A, M)   & \cmark (D, E)     \\
    \hline
    Crossroads with/without obstacle (5.11) &  &  & \cmark & \cmark & \cmark (S)  & \cmark (A,  M) &\cmark (E)    \\
    \hline
    UAV and robot collision avoidance (5.12, 5.23) & \cmark & \cmark & \cmark &   &  \cmark (D, W) & \cmark (R)  &  \cmark (D,  E) \\
\hline
    Tourist spot traffic management  (5.14) & 
    \cmark &  & \cmark & \cmark  & \cmark (D, W) &  \cmark (A) &      \\
    \hline
    Sleep and health monitoring  (5.15, 5.17, 5.18) & \cmark & \cmark &  &   & \cmark(D, W)  & \cmark (A) & \cmark (D)    \\
    \hline
    Sensor groups (5.19) &  &  & \cmark & \cmark  &  \cmark (D, S) & \cmark (R, A, M) & \cmark (D, E)   \\
    \hline
     Parking space determination (5.20) & 
     \cmark & \cmark &  &   &   \cmark (W) & \cmark (R, A) & \cmark (E)     \\
    \hline
    Seamless XR streaming (5.21) & 
     &  & \cmark & \cmark &  \cmark (S) & \cmark (R, A, M)  &     \\
    \hline
    Immersive experience  (5.25) & 
     &  & \cmark & \cmark  &  & \cmark (A) & \cmark (E)     \\
    \hline
    Public safety (5.27) &  
    \cmark & \cmark & \cmark &  & \cmark (D,  W, S)  & \cmark (M)  &  \cmark (D,  E)  \\
    \hline
    Vehicles sensing for ADAS (5.28) &  
    \cmark &  & \cmark & \cmark & \cmark (D, S)  &  & \cmark (E)   \\
    \hline
    Gesture recognition (5.29, 5.24)  & 
    \cmark &  & \cmark & \cmark &   & \cmark (A)  & \cmark  (E)   \\
    \hline
    Blind spot detection (5.31) & 
    \cmark &  & \cmark & \cmark &  \cmark (A) &  &      \\
    \hline
    Sensing and positioning in factory hall (5.32) &
    \cmark & \cmark & \cmark & \cmark  & \cmark (W, S) & \cmark (R, A, M) &  \cmark (D, E)     \\
    \hline
    \end{tabular}
                    }}
     \vspace{1mm}
    \caption{{An overview of the \ac{ISAC} use cases from \cite[Section 5]{3GPPTR22837}, the four 6G services from \cite[Section 2.2]{CalvaneseGOWSC2021} 
    (i.e., 
 massive machine-type communications supporting distributed intelligence (MMTCCxDI); globally-enhanced mobile broadband (GeMBB);  ultra-reliable, low-latency computation, communication, and control (URLLCCC);  semantic services) 
    and whether and how they would benefit from the DISAC cornerstones summarized in Fig.~\ref{fig:DISAC-section1} (if not, the cell is empty). ADAS: advanced driver assistance system; AGV: automated guided vehicle; UAV: unmanned aerial vehicle; XR: extended reality. }}

    \label{tab:UseCases}
    \vspace{-8mm}
\end{table*}

{Among the many demanding use cases with a wide range of requirements considered by 6G, one that stands out for \ac{ISAC} are digital twins~\cite{Masaracchia_DT2024}, as it subsumes many other challenging use cases from \cite{3GPPTR22837} as special cases (e.g., intrusion detection, navigation, tracking, collision avoidance, and public safety).}
A digital twin
is a digital representation of a possibly large-scale physical system (including factories, warehouses, but also buildings, or even entire cities) on a computer or a cloud-based platform, to continuously monitor its status and predict future states, thus, facilitating the system's maintenance and relevant decision making. 
{An example is the \emph{urban digital twin}, a digital/virtual representation of a city made possible/enriched by vehicles equipped with a diverse array of sophisticated sensors, encompassing radar, lidar, and cameras, which results in the creation of a dynamic, mobile, and distributed 6G user-centric sensor network on roadways. This representation is complemented by sensors connected to the 5G/6G urban infrastructure and non-3GPP sensors, such as traffic cameras, underground road sensors, and weather stations, resulting in a large-scale sensor network spanning the same urban environment.}
The seamless exchange of information between the various sensors and the network core, as well as the 
symbiotic interplay of distributed sensing and communication is enabled by the DISAC concept.

\subsection{The Standardization Push}
Several \acp{SDO} have begun to explore and develop work on the ISAC topic in recent years, by integrating \ac{AI}-based approaches with varying degrees of progress. We summarize the objectives and status of the main \acp{SDO} below, also visualized in Fig.~\ref{fig:std_RM}.

 \subsubsection{Progress in \ac{ISAC}}

In 2023's last quarter, {ETSI} established a new industry specification group (ISG) dedicated to \ac{ISAC}~\cite{ETSIISG}. The purpose of this initiative is to complement the ongoing efforts of 3GPP, with a specific focus on channel modeling in Release 19,
and to address complementary subjects that are ahead of the current 3GPP scope, including, e.g., radio access network (RAN) and system architecture, as well as privacy, security, and trustworthiness of sensing data. DISAC has direct implications to the RAN and architecture aspects, which should be incorporated during 2024-2025.
Second, {3GPP} completed a feasibility study on \ac{ISAC} use cases and requirements \cite{3GPPTR22837} and started the specification work \cite{3GPPTS22137} covering sensing operation and functional requirements, including security and secrecy aspects, as well as performance requirements. More than 30 use cases have been agreed, covering broad applications at home, industry, and city. A study item focusing on channel modeling will start in RAN1 in Spring 2024, to be possibly extended to additional topics by the end of 2024, depending on the work progress. The consistency of such models over extended spatial and temporal domains is of direct importance to DISAC. 
Third, {ITU-R WP5D}  will consider ISAC as a new usage scenario to enable innovative services and solutions~\cite{ITUR2160}. Communications can assist sensing services (``network as a sensor''), and sensing can be used to assist communications as a new channel that links the physical and the digital worlds (a.k.a. sensing-aided communications). The semantic nature of DISAC is expected to play a pivotal role here since the corresponding context awareness will significantly reduce overheads, control, and signaling. 
{Beyond the SDOs in cellular systems, IEEE 802.11 has launched the 802.11bf Task Group ~\cite{IEEE80211bf} already in 2020. This amendment defined a few modifications in the \ac{PHY} and \ac{MAC} layers to enhance sensing operations of \acp{WLAN} in unlicensed frequency bands above 45 GHz,}
with applications that include presence detection, target tracking, gesture classification, gesture control, imaging, and vital signs monitoring. 802.11bf can be used to obtain channel measurements that characterize the environment in which the sensing stations (either \acp{AP} or client devices) operate, supporting monostatic, bistatic, and multi-static sensing.
\begin{figure*}
    \centering
    \includegraphics[width=1.5\columnwidth]{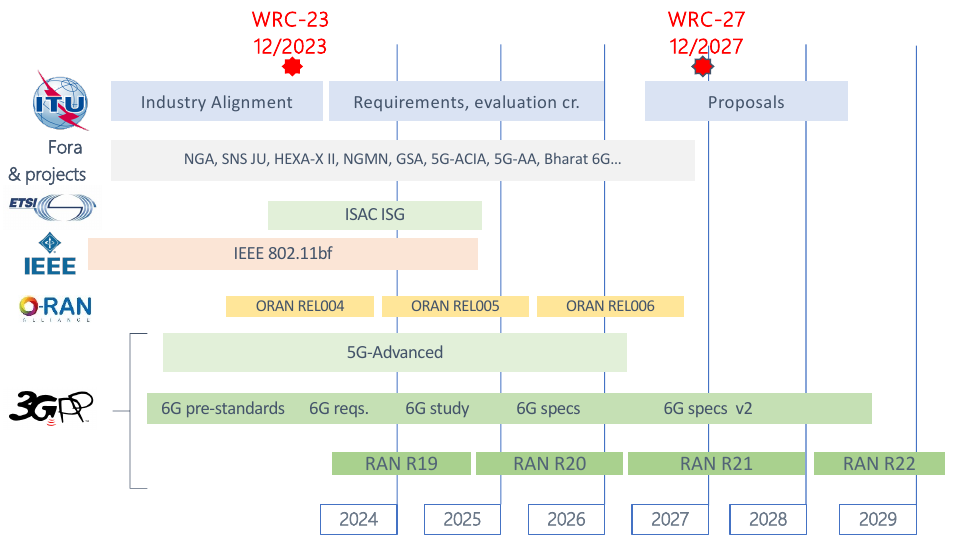}
    \caption{The standardization roadmap covering different \acp{SDO}.}
    \label{fig:std_RM}\vspace{-5mm}
\end{figure*}

\subsubsection{Progress on AI/ML}
Feasibility studies and normative work on AI/ML have already begun at 3GPP and IEEE. In 3GPP, a normative work has been approved for Release 18, focusing on enhancements to data collection and signaling to support AI/ML-based network energy savings, load balancing, and mobility optimizations. This work continues in Release 19 to further enable AI/ML for the air interface and core network, including efficient training, data management, and trustworthiness AI/ML. The support of ISAC for AI/ML will start when feasibility studies have progressed sufficiently, probably in 3GPP Release 20. 
{IEEE completed an AI/ML technical interest group (TIG) in IEEE 802.11 in March 2024 and has formed a Standing Committee  having in scope the description of AI/ML use cases, the study of their applicability to 802.11 systems, as well as their technical feasibility. Exploration of new AI/ML-enabled applications in 802.11 systems are also in scope, which may include ISAC.}

Critically, ISAC-based support is not yet included in the aforementioned AI/ML-related frameworks, but it will benefit from the standardization methodology developed for other applications for fast introduction. Due to the inherent semantic nature of sensing information, the DISAC vision directly implies tight coordination between ongoing ISAC and AI/ML standardization efforts. 

\section{The DISAC Technological Enablers}

{The DISAC vision described sofar relies on four distinct enablers: \textit{i)} an overarching semantic framework for ISAC (related to the semantic approach cornerstone from Fig.~\ref{fig:DISAC-section1}); \textit{ii)} an optimized and parsimonious physical layer based on a dedicated signal design and high-resolution processing for sensing, complemented with sensing-aided communications; \textit{iii)} corresponding intelligent resource allocation (both related to the high-resolution processing cornerstone from Fig.~\ref{fig:DISAC-section1}); and \textit{iv)} an evolved architecture for supporting the previous enablers (related to the DISAC architecture cornerstone from Fig.~\ref{fig:DISAC-section1}).}

\subsection{Native Semantic Framework}
\label{sec:EnablersSEM}
DISAC capitalizes on the semantic and goal-oriented communications approach in~\cite{CalvaneseGOWSC2021} as a crucial component. Including semantic and goal-oriented aspects into \ac{ISAC} allows diverse information from various sensing modalities to be aggregated, facilitating the transition from traditional data fusion to the composition of semantically selected information and the pragmatic generation of AI-based reasoning stimuli. These implications extend to all aspects of DISAC, including signal shaping, resource allocation, signal processing, computation, storage, and architecture.
 
Leveraging semantic and goal-oriented communications for distributed ISAC provides advantages in terms of interoperability, contextual understanding, and effectiveness in activating sensing functions. This includes context- and application-adapted waveform design, signal processing, dedicated resource allocation, robust protocols, as well as reasoning about multi-modal sensed information. We propose to establish and exploit connections or relationships between different pieces of information derived from sensed and already available data, based on causal and semantic representation, extraction, communications, and reasoning-based interpretation and composition. This unified understanding is instrumental to effectively analyze, interpret, and derive insights from data, thus optimizing the use of sensing, computation, and communication resources.

\begin{figure}
    \centering    
    \includegraphics[width=0.9\columnwidth]{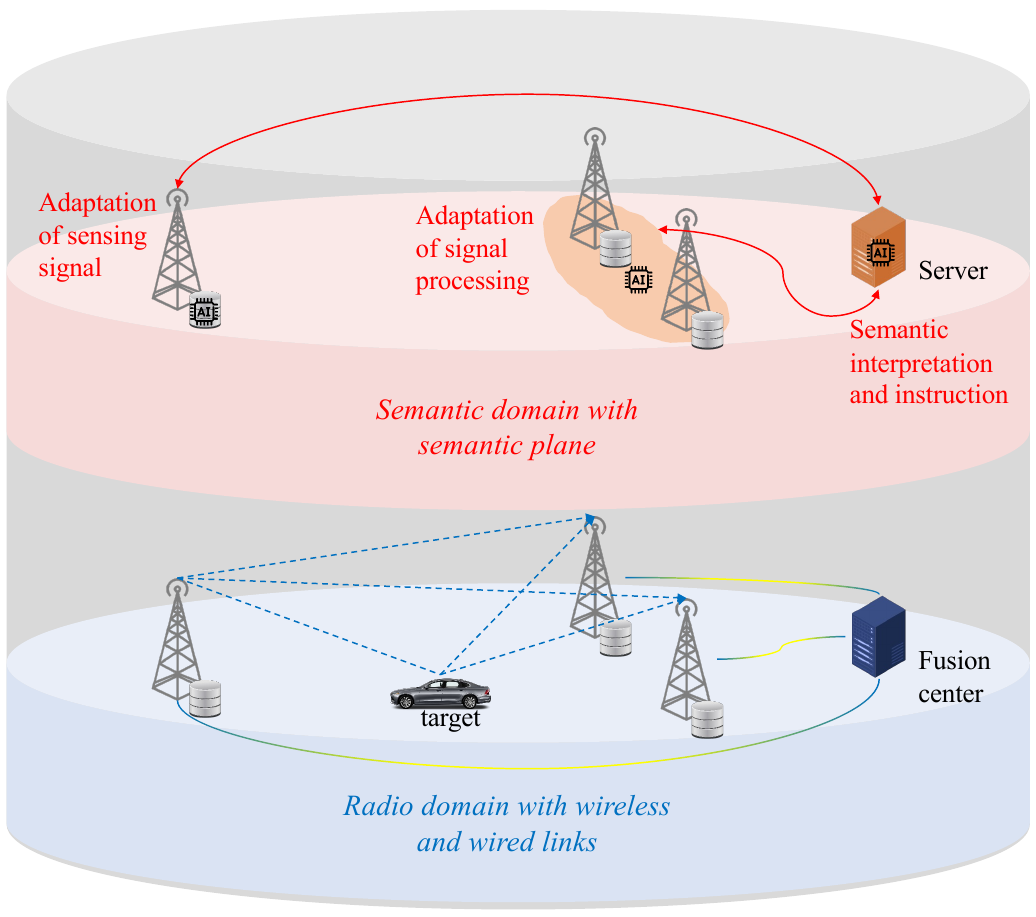}
    \caption{The radio domain ({bottom}) and the DISAC semantic domain ({top}), which provides improved robustness and resource allocation efficiency, for an example multi-static sensing scenario.}
    \label{fig:semanticDomain}
\end{figure}

The relation between the radio and the semantic domains is visualized in Fig.~\ref{fig:semanticDomain}, considering a multi-static sensing scenario as an example. The semantic channel between the server running the application and the base stations, responsible for transmission/reception and processing, supports intelligent control on what should be transmitted on the radio domain, how the signal should be processed, and which information should be sent to the server, again in the radio domain. Over time, AI is expected to learn a more parsimonious and efficient information representation, leading to increasing performance gains (e.g., less power consumption and fewer overheads).

\subsection{Optimized and Parsimonious Physical Layer}

\label{sec:EnablersDSP}

{The DISAC physical layer involves components of classical signal processing, each of which must be extended to a distributed context. These components include signal optimization (i.e., optimizing resource allocation across time, frequency, and spatial dimensions); channel parameter estimation to extracting position-related parameters, such as ranges, angles, and Dopplers from received waveforms; detection and estimation of objects and connected devices; and tracking of \acp{UE} and objects over time, including solving data association problems. There has been research on distributed versions in each of these areas for over the last 15 years, to some extent within the realm of communications, but more often in (distributed) radar processing and multi-robot interaction. Nevertheless, as pointed out in \cite{tong2023multi}, extending classical signal processing principles to the DISAC context is challenging when considering the DISAC vision (see Fig.~\ref{fig:DISAC-section1}), which in turn motivates the study of new enablers and methods at the physical layer.}

\subsubsection{Semantic Transceiver Processing}
Moving away from the perspective of optimizing \ac{ISAC} \acp{KPI}, a semantic approach requires revisiting waveform optimization, channel parameter estimation, location estimation, and tracking to tailor to the instantaneous application needs. This completely changes the optimization and learning approaches and, if executed correctly, will lead to a more parsimonious \ac{ISAC} operation. Nevertheless, to reach its full potential, including semantics as a cornerstone of DISAC, requires a dedicated semantic plane, as illustrated in Fig.~\ref{fig:architecture}.
In terms of signal shaping, the use of semantic-aware encoding/decoding over multiple frequency bands can enhance the performance and robustness of sensing and communications, while flexible coordination among transmitted waveforms (e.g., based on OFDM or even learned from data) can provide improved efficiency, resilience, and robustness against interference and hardware distortion {\cite[Section 4.2]{CalvaneseGOWSC2021}}. 
{For the receiver-side processing, 
semantic processing for efficient data representation is foreseen in combination with \ac{AI}/\ac{ML} (e.g., for imaging and extended target processing) and model-based methods (e.g., for deciding which sensors to activate and when). For this, Bayesian inference and concepts from model-based control will be useful tools.}

\subsubsection{Distributed and High-Resolution Sensing}
Resolution is required for all sensing applications, as it determines the ability to separate and distinguish objects. Resolution comes in different and mutually compounding forms, such as delay resolution (with large bandwidths), Doppler resolution (with long integration times), and angle resolution (with large antenna sizes). In DISAC, an additional dimension is available, namely the distributed aperture. An example of the inherent aperture resolution provided by a distributed system is visualized in Fig.~\ref{fig:quantitative-results} via the ambiguity function. From the distributed aperture alone, high resolution (on the order of the wavelength) is attained, while sidelobes are suppressed by the random placement of the many distributed receivers \cite{lehmann2006high}. Realizing and harnessing this resolution is challenging. First of all, high accuracy requires extreme calibration of the infrastructure, including the static aspect (the locations of \acp{AP}) and the dynamic aspect (time, frequency, and phase synchronization). %The latter part demands new over-the-air methods to ensure coherence between the observations at different sites. 
Secondly, depending on the level of coherence/synchronization between different \acp{AP}, different high-resolution methods should be applied and different amounts of data should be shared: phase-coherent processing requires sharing of raw I/Q data, while time-coherent processing requires sharing of estimated channel parameters. This problem is further exacerbated by the different fields of view of different nodes so that each \ac{AP} sees the same environment from a different perspective. Fusing these perspectives brings the most informative picture of the environment, but it is also the most challenging one. Methods from distributed \ac{MIMO} radar and imaging in synthetic aperture radar are expected to play an important role. {Finally, the processing must account for the large diversity of \ac{MIMO} technologies, such as \acp{RIS} (providing controlled signal paths), \ac{D-MIMO} (providing user-centric multi-point connectivity), and \ac{XL-MIMO} (providing large apertures with beneficial near-field propagation effects), as well as their hardware models and impairments.}

\subsubsection{Sensing-Aided Communications}
\label{sec:EnablersCAS}
\begin{figure*}
    \centering
    \includegraphics[width=1.5\columnwidth]{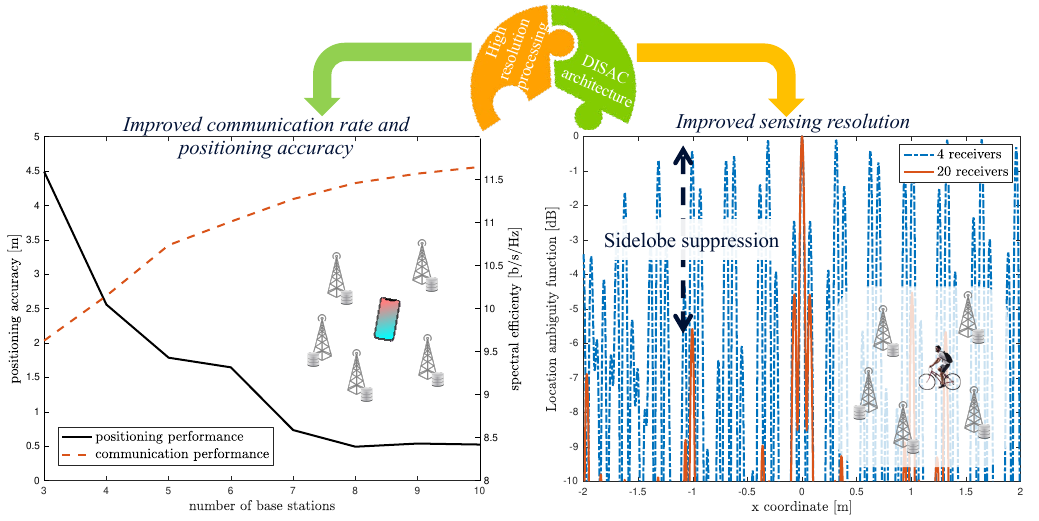}
    \caption{Examples of quantifiable performance gains of DISAC. {Left: Experimental evaluation of a time-coherent D-MIMO system at 2.35 GHz in terms of positioning and communication \acp{KPI}
    \cite{Loc_DMIMO_2022}.}
    {Right: The ambiguity function computed from the  near-field steering vector of 4 and 20 phase-coherent receivers (randomly placed in a 1D environment $[-50~\text{m},+50~\text{m}]$) for a single-tone signal at 3 GHz. More receivers yield benefits in terms of inherent resolution and low sidelobes \cite{lehmann2006high}.}}
    \label{fig:quantitative-results}
\end{figure*}

A final component of the physical layer, and the associated signal processing, is the exploitation of sensing and contextual information to improve communication operations.
The optimization of the wireless communication system's components requires, in general, the estimation of the parameters of the channel conditions in which it operates. Conventionally, the channel estimation is based on the exchange of pilot signals in time, frequency, and space, providing a snapshot of the channel. 
Such estimation processes need to be efficiently supported by the communication protocol with dedicated control phases and signaling, aiming for accurate, low latency, and scalable estimation. 
This classical approach becomes very challenging in dense smart wireless environments designed to use XL-\ac{MIMO} nodes and multiple \acp{RIS}~\cite{RISE6G_COMMAG}, possibly realizing either hybrid analog and digital or fully digital beamforming with very low-resolution signal converters and passive reflective beamforming, especially in very high-frequency bands and under mobility scenarios. The DISAC paradigm envisions the optimization of large-scale and distributed communication provisioning with location and, more generally, context information becoming available from the distributed processing of data gathered from multi-modal sensory devices integrated within the wireless network infrastructure, paving the way for more efficient, context-aware, and environmentally sustainable communication networks. {Sensing-aided communication has been shown to offer significant performance gains, especially for full-duplex MIMO \cite{Smi25}.
The inherent synergy between positioning and communication performance is shown in Fig.~\ref{fig:quantitative-results}, supporting the fact the D-MIMO systems increase both communication and positioning \acp{KPI}.} 

\subsection{Intelligent Resource Allocation}
\label{sec:EnablersRA}

{The development of adaptive resource allocation schemes is fundamental within the DISAC concept. Such techniques need to further enable simultaneous sensing and communications in dynamically evolving scenarios with heterogeneous nodes. Unlike traditional resource allocation strategies, which primarily focus on bandwidth and power constraints to provide communication services, DISAC-oriented resource allocation policies must account for the unique demands of high-resolution sensing (leading to potentially large volumes of sensor data) and semantic processing (which will reduce these data volumes intelligently over time and space) to effectively face the distributed nature of the heterogeneous nodes DISAC foundationally relies on.}

\subsubsection{Goal-Oriented and Context Awareness} Efficient resource allocation schemes must be inherently goal-oriented, exploiting contextual information to strike a balance in the time, frequency, and energy resources allocated for sensing and communications {among distributed nodes}. This dynamic adjustment to varying network conditions and user requirements is crucial, especially considering the heterogeneous nature of future wireless networks, where devices range from low-powered Internet-of-Things (IoT) sensors to high-capacity data centers. Interestingly, incorporating advanced signal processing methods and the novel semantic domain (see Section~\ref{sec:EnablersSEM}) within DISAC, the resulting adaptive resource allocation strategies will ensure that both sensing data and communication signals are processed and transmitted efficiently. This does not only address the integrity or the timeliness of the data but also the relevance of the content, prioritizing critical information for respective transmission.

\subsubsection{XL-MIMO and Smart Wireless Environments}
Adaptive resource allocation plays a vital role in harmonizing the sensing and communication components, and ensuring their seamless operation and synergy when considering emerging multi-antenna technologies, such as XL-MIMO (co-located or D-MIMO) and multi-functional \acp{RIS}. Transceivers with extremely large arrays of antennas or metamaterials offer design flexibility for both far- and near-field DISAC, necessitating, however, efficient utilization of the spatiotemporal, frequency, and power resources to achieve high spectral/energy efficiencies and sensing accuracy. The same holds for D-MIMO deployments, with possibly heterogeneous multi-antenna technologies, where efficient coordination schemes are mandatory. In addition, the DISAC vision incorporates smart wireless connectivity schemes enabled by the \ac{RIS} technology~\cite{RISE6G_COMMAG}, which offers a power- and cost-efficient scalable solution for distributed communications as well as sensing of wide perspective over space. \ac{RIS}-enabled smart wireless environments necessitate advanced resource allocation schemes that may reach the level of managing tiles among different \acp{RIS} for efficient ISAC. To this end, multi-functional \acp{RIS} may play a profound role in realizing DISAC's radio and semantic domains.

\subsection{An Evolved Network Architecture} 
\label{sec:EnablersArch}  
\begin{figure}
    \centering
    \includegraphics[width=0.99\columnwidth]{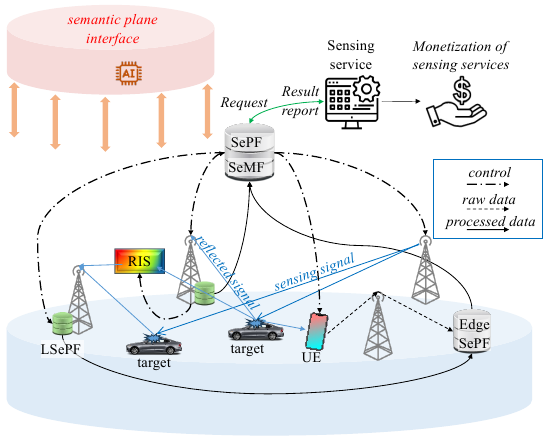}
    \caption{{The core components of the DISAC network architecture.} }% with involved \ac{SRN} and \ac{LDP} items.}
    \label{fig:architecture}
\end{figure}

The DISAC architecture, whose core components are illustrated in Fig.~\ref{fig:architecture}, will depart from the conventional cellular network architecture in the following ways.

\subsubsection{Novel Network Functions and Interfaces}
{The 3GPP network architecture offers support for various communication requirements (eMBB, URLLC, and mMTC) and the positioning of UEs. Although the architectural components enabling positioning provide a basis for sensing, significant changes are needed to fully integrate sensing in the architecture. Moreover, sensing needs to be integrated within 3GPP-compliant systems, that are communications-wise optimized, to provide seamless communications and sensing using the same infrastructure and resources. Complementing the envisioned sensing management function (SeMF), the DISAC architecture provides for sensing signal processing local to the node (LSePF), at the edge, or at the core (SePF).
%To this end, the DISAC architecture needs to support distributed intelligent signal processing at the \acp{SRN}, so that only higher-level sensing information is sent to the fusion center. 
This involves defining new network functions, protocols, and interfaces. In addition, new functions dedicated to tracking and handover of objects are necessary. This in turn requires the definition and management of node groups, as well as suitable identifiers for objects that do not possess subscriber identity module (SIM) cards. Finally, existing interfaces to external sensors (e.g., GNSS) must be augmented to account for a wider variety of sensors (e.g., cameras and lidar).}

{\subsubsection{Distributed Processing with Heterogeneous Devices}
The DISAC architecture aims to detect and track multiple targets using multiple sensing nodes over a geographical area of interest for different applications. This involves many transmitting and receiving nodes, as well as multiple \acp{RIS}, each with limitations in terms of, e.g., bandwidth, power, processing capability (including \ac{MEC}), or backhauling constraints. Critical challenges are the optimal selection, representation, and compression of exchanged data to enable scalable collaborative fusion across different distributed sensing nodes
%In this context, a critical challenge to be faced is the optimization of the amount of data to be exchanged 
%to perform the fusion involving distributed and collaborative processing among the different sensing nodes
~\cite{Klupacs2022multiagent}.  The architecture must thus support a variety of heterogeneous sensing devices (e.g., multi-functional RISs and MIMO), each with its capabilities and limitations in terms of, e.g., bandwidth, power, processing capability, or backhauling constraints. Hence, the consideration of heterogeneous energy budgets,  computation constraints, and data representations
of different types of nodes is crucial in the design of distributed optimization algorithms to balance the load, based on the instructions from the semantic layer, to optimally perform local processing and hierarchical fusion of information. In general, the architecture needs to support distributed functions across multiple nodes and planes. These functions include synchronization and calibration, which can be implemented using either wired or wireless control communication connections. %The integration of multi-functional RISs that can be used both for improving communications and sensing, for assisting in network deployment optimization, and for designing resource allocation and orchestration solutions constitute critical DISAC aspects. 
}

\subsubsection{Semantic Layer}
{The inclusion of the semantic / goal-oriented plane for closed-loop representation and control in DISAC operation will enable the adaptation of the ISAC  parameters (refresh rate, performance criteria) and resource allocation to varying KPI/KVI requirements. It will also support other 6G services beyond DISAC.} 
DISAC envisions semantic information exchange between different network functions, which implies the introduction of new protocols and network functions that include semantic extraction, semantic composition, and semantic instruction. These functions must also interact with the radio layer, through operator functions that typically control network nodes. Among the distributed functions that need to be supported by the DISAC network architecture, the relevant AI modules at the semantic layer but also those at the radio layer stand out, while for training, inference, and exchange of information to enable federated learning and multi-agent learning, respective efficient functions and interfaces need to be designed. 

%\subsubsection{XL-MIMO- and RIS-Aided Sensing}
%\subsubsection{Support for Heterogeneous Devices}
%DISAC aims to support a wide diversity of sensing devices, and related networks elements involving different sensing modalities and communication patterns, each with its capabilities and limitations in terms of, e.g., bandwidth, power, processing capability, or backhauling constraints. Hence, the consideration of heterogeneous energy budgets and computation constraints of different types of nodes is crucial in the design of distributed optimization algorithms for sensing and communications. 
%While the current 3GPP architecture supports MIMO, different types of RISs are expected to require additional considerations. Passive RISs can be transparent from the architecture viewpoint, but they still need to be efficiently managed. In addition, RISs with signal reception capabilities can act as \acp{SRN}, but with limited local processing capabilities. 
%The design and integration of multi-functional RISs that can be used both for improving communications and sensing, for assisting in network deployment optimization, and for designing resource allocation and orchestration solutions constitute critical DISAC aspects. 

\subsubsection{Business Potential}
Leveraging the large number of 3GPP-compliant installed bases on a global scale, network sensing can provide cost, scale, and coverage advantages over existing non-3GPP-based sensing solutions. This creates opportunities for \acp{CSP} to monetize the network infrastructure with additional data collection and new value-added services. Thus, distributed sensing offers valuable service differentiation opportunities for operators and the network applications ecosystem.

\section{Challenges in Realizing the DISAC Vision}
The DISAC vision is expected to increase the TRL of ISAC. By seamlessly integrating sensing with communications and harnessing the power of distributed AI, \ac{DISAC} not only addresses the limitations of existing \ac{ISAC} approaches, but also unlocks new possibilities for resource-efficient, accurate, and semantic network operations. While many of DISAC's building blocks (e.g., semantic communications, distributed computing, high-resolution sensing, and sensing-aided communications) have been and are being studied in isolation, bringing them together requires facing challenges that can be tackled with a broad view, encompassing stakeholders from industry and academia, covering all aspects of the 6G value chain. Specific challenges can be broken down as follows. 

\subsubsection{Theory and Algorithms}
First of all, designing and executing the semantic framework  presents a formidable challenge. Despite its enormous promise and potential, it significantly departs from the standard approach and architecture for ISAC and it will, thus, need extensive study in the coming years. Secondly, the distributed approach for ISAC will require new ways of shaping and processing the wireless signal, and fusing the resulting information with external sensors, as well as considering this information over extended space and time. Raw computational and communication constraints are expected to be the main bottlenecks, thus, requiring lightweight processing methods.

\subsubsection{Proofs of Concept}
Current sensing demonstrations predominantly showcase \ac{ISAC} in a non-distributed paradigm using standardized waveforms, where the ongoing investigations predominantly take place at lower layers of the communication stack~\cite{Zhang2023Networking}. In contrast, DISAC considers not only optimized waveforms but also the fusion of sensory information from various sensors, which all lead to traffic over the communication network. These sensors include the distributed 6G-compliant nodes, as well as external sensors. The provided information must be compressed, synchronized, and fused, either outside the network or within respective core functionalities. One of the challenges is the development of new hardware components that can operate in different frequency ranges while supporting ISAC in distributed settings. In this context, hardware components must be designed to work together seamlessly and they need to be able to provide interfaces to existing systems. Furthermore, the use of multi-functional \acp{RIS} adds another layer of complexity, especially if the \acp{RIS} cannot sense locally. An additional challenge relates to the real-world demonstration of the semantic framework. Even a simple single-link semantic solution to sensing is challenging, given the need for training data and the lack of theory on semantic communications in the context of ISAC. Further research challenges include the implementation of algorithms and protocols for sensory data sharing, new signal and information processing techniques, and the optimization of the system performance in different scenarios in real-world demonstration platforms.

\subsubsection{Standardization}
While some sensing use cases do not require standardization (e.g., network-based sensing without \acp{UE} or \ac{UE}-based sensing without base stations), many rely on both the \ac{UE} and the network infrastructure and, thus, require standardization. Studies on waveforms and distributed architectures, including the definition of new functional elements, protocols, and interfaces are needed for discussion in \acp{SDO}. Furthermore, specific metrics, KPIs, and KVIs applicable to DISAC must be developed, as well as suitable channel models, to evaluate performance in the relevant use cases. Finally, coordination between ISAC and AI activities in \acp{SDO} is needed as well, in support of the semantic framework.

\bibliographystyle{IEEEtran}
\bibliography{references}

\begin{IEEEbiographynophoto}{Emilio Calvanese Strinati} is the 6G Future Technologies Director at the French Atomic Energy Commission's Electronics and Information Technologies Laboratory, Minatec Campus, Grenoble, France with research interest in semantic communications and ISAC. He is the coordinator of 6G-DISAC and 6G-GOALS SNS-JU-funded projects.
\end{IEEEbiographynophoto}
\vskip -2\baselineskip plus -1fill

\begin{IEEEbiographynophoto}{George C. Alexandropoulos} 
is an Associate Professor in the Department of Informatics and Telecommunications, National and Kapodistrian University
of Athens, Greece  and an Adjunct Professor with the Department of Electrical and Computer Engineering, University of Illinois Chicago, USA. His research interests span the general areas of algorithmic design and performance analysis for wireless networks with emphasis on multi-antenna transceiver hardware architectures, full duplex MIMO, active and passive multi-functional RISs, ISAC, and millimeter-wave/THz
communications, as well as distributed machine learning algorithms. 
%is an Associate Professor in the Department of Informatics and Telecommunications, National and Kapodistrian University of Athens, 15784 Athens, Greece. His research interests span the general areas of algorithmic design and performance analysis for wireless networks with emphasis on multi-antenna transceiver hardware architectures, full duplex MIMO, active and passive multi-functional reconfigurable intelligent surfaces, integrated sensing and communications, millimeter-wave/THz communications, as well as distributed machine learning algorithms. He has participated and/or technically managed numerous EU, international, and Greek research, innovation, and development projects, including H2020 RISE-6G, ESA PRISM, SNS JU TERRAMETA, and SNS JU 6G-DISAC.
\end{IEEEbiographynophoto}
\vskip -2\baselineskip plus -1fill

\begin{IEEEbiographynophoto}{Navid Amani}  
 received his M.Sc. degree in Telecommunication Engineering from K. N. Toosi University of Technology, Tehran, Iran, in 2013. In 2021, he obtained the Ph.D. degree from Chalmers University of Technology, Gothenburg, Sweden and  from Eindhoven University of Technology, Eindhoven, The Netherlands. He was a Visiting Researcher at the Netherlands Institute for Radio Astronomy and NXP Semiconductors, Eindhoven. He is the founder of EMickers AB and also a researcher at Chalmers University of Technology, both based in Gothenburg, Sweden. His research interests include distributed MIMO systems and low-power IoT devices.
%received the M.Sc. degree in telecommunication engineering from K. N. Toosi University of Technology, Tehran, Iran, in 2013, and the Ph.D. degree from Chalmers University of Technology, Gothenburg, Sweden, in 2021, as a Marie Sk{\l}odowska-Curie Actions (MSCA) Fellow. He also obtained a double Ph.D. degree from the Electromagnetics Group, Eindhoven University of Technology (TU/e), Eindhoven, The Netherlands, in 2021. He is a co-founder of Radchat AB in Gothenburg, Sweden, where he leads the technical development of radar-centric joint communication and sensing systems.
\end{IEEEbiographynophoto}
\vskip -2\baselineskip plus -1fill

\begin{IEEEbiographynophoto}{Maurizio Crozzoli} is an engineer at TIM in the Wireless Access Innovation department. He worked on the development and testing of Active Antenna Systems (AAS) and his recent research activities include Reconfigurable Intelligent Surfaces (RIS).
\end{IEEEbiographynophoto}
\vskip -2\baselineskip plus -1fill

\begin{IEEEbiographynophoto}{Giyyarpuram Madhusudan} is a senior research engineer at Orange Labs. He leads the research activity on 5G/6G Sensing at Orange. He represents Orange at the ETSI ISG on ISAC. His research interests include network technologies and architecture for IoT.
\end{IEEEbiographynophoto}
\vskip -2\baselineskip plus -1fill

\begin{IEEEbiographynophoto}{Sami Mekki} received the Engineering Diploma degree in wireless networks from SUP'COM, Tunis, Tunisia, in 2004, the M.Sc. degree in signal and digital communication from the University of Nice Sophia-Antipolis, in 2005, and the Ph.D. degree in electrical engineering from Telecom ParisTech, in 2009. He has more than 15 years of experience in the industrial research domain, where he has published several papers and patents in various fields. Since 2022, Sami has joined Nokia Networks France as a Senior Radio Research Specialist working in the Standardization department. His research interests include wireless communication, channel estimation, multiuser detection, reconfigurable intelligent surfaces, automotive driving, and sensor data fusion for user localization.

\end{IEEEbiographynophoto}
\vskip -2\baselineskip plus -1fill

\begin{IEEEbiographynophoto}{Francois Rivet} is an Associate Professor at the EE department of the Bordeaux Institute of Technology and IMS laboratory, in Bordeaux, France. His research activities are the design of integrated circuits and systems for wireless communications. Since 2014, he has led the research team 'Circuits and Systems'. He has contributed to the design of disruptive communication circuits developing a Design by Mathematics methodology. He published 150 technical papers and holds 19 patents.
\end{IEEEbiographynophoto}
\vskip -2\baselineskip plus -1fill

\begin{IEEEbiographynophoto}{Vincenzo Sciancalepore} received his M.Sc. degree in telecommunications engineering and telematics engineering in 2011 and 2012, respectively, and in 2015, he received a double Ph.D. degree. Currently, he is a Principal Researcher at NEC Laboratories Europe, Germany, focusing his activity on RISs. He is an editor of IEEE Transactions on Wireless Communications and IEEE Transactions on Communications. 
\end{IEEEbiographynophoto}
\vskip -2\baselineskip plus -1fill

\begin{IEEEbiographynophoto}{Philippe Sehier}  Philippe Sehier is a principal research lead Nokia Strategy and Technology in Nokia France. He received his engineer degree from {Ecole Sup\'{e}rieure D'\'{e}lectricit\'{e} (Supelec)} in 1984. His current research areas are focused on the 5G and 6G radio access in close relationship with 3GPP standardization. Before this position, he was a delegation head in several standardization bodies. He also occupied several positions in spatial and military activities for Alcatel-Lucent. 
\end{IEEEbiographynophoto}
\vskip -2\baselineskip plus -1fill

\begin{IEEEbiographynophoto}{Maximilian Stark} 
received the B.Sc., M.Sc., and Ph.D. degrees (Hons.) in electrical engineering from the Hamburg University of Technology (TUHH) in 2014, 2017, and 2021, respectively. In 2019, he was with Nokia Bell Labs, France, as a Visiting Researcher in the area of deep learning in communications. During 2021--2024, he joined Bosch Research, where he was responsible for the development of AI/ML methods for V2X communications and reliable positioning. Currently, he is with NXP Semiconductors in Germany, working as a Principal System Engineer in the CTO Office for System Innovation. %His work focuses on 6G research, specifically efficient baseband signal processing and its integration with AI. He was a recipient of the Karl H. Dietze Award in 2017 for his master’s thesis, the Sick Award in 2022, and the Tesat-Spacecom Science Award in 2023 for his Ph.D. thesis.
%
%He was a recipient of the Karl H. Dietze Award in 2017 for his master's thesis, the Sick Award in 2022, and the Tesat-Spacecom Science Award in 2023 for his Ph.D. Thesis. 
\end{IEEEbiographynophoto}
\vskip -2\baselineskip plus -1fill

\begin{IEEEbiographynophoto}{Henk Wymeersch}  is a Professor at Chalmers University of Technology, Sweden, active in the area of radio localization and sensing. He has been active in several European and Swedish industry-academic projects on ISAC and is also vice-chair of the ETSI  industry specification group on ISAC. 
\end{IEEEbiographynophoto}
\balance 
\end{document}

%% file: acronyms.tex
\usepackage[nolist]{acronym}
\begin{acronym}[ACRONYM]
\acro{3GPP}{The 3rd Generation Partnership Project}
\acro{3D}{three dimensional}
\acro{5G}{fifth generation}
\acro{6G}{sixth generation}
\acro{AI}{artificial intelligence}
\acro{AP}{access point}
\acro{AR}{augmented reality}
\acro{AoA}{angle-of-arrival}
\acro{AoD}{angle-of-departure}
\acro{BS}{base station}
\acro{CFO}{carrier frequency offset}
\acro{CNN}{convolutional neural network}
\acro{CSI}{channel state information}
\acro{CSP}{communication service provider}
\acro{CU}{central unit}
\acro{DL}{downlink}
\acro{DoF}{degrees of freedom}
\acro{DU}{distributed unit}
\acro{D-MIMO}{distributed MIMO}
\acro{DISAC}{distributed and intelligent integrated sensing and communications}
\acro{DT}{digital twin}
\acro{EER}{energy efficiency ratio}
\acro{ELAA}{electrically large aperture array}
\acro{EM}{electromagnetic}
\acro{FD}{Full Duplex}
\acro{GDOP}{geometric dilution of precision}
\acro{GNSS}{global navigation satellite system}
\acro{HMI}{human-machine interaction}
\acro{IQ}{in-phase and quadrature}
\acro{IP}{incidence point}
\acro{JCS}{joint communications and sensing}
\acro{ISAC}{integrated sensing and communications}
\acro{JRC}{joint radar and communication}
\acro{JRC2LS}{joint radar communication, computation, localization, and sensing}
\acro{ICI}{inter-carrier interference}
\acro{IOO}{indoor open office}
\acro{IoT}{Internet of Things}
\acro{IRN}{infrastructure reference node}
\acro{ISG}{industry specification group}
\acro{JCAS}{joint communications and sensing}
\acro{KPI}{key performance indicator}
\acro{KVI}{key value indicator}
\acro{LCA}{life cycle analysis}
\acrodefplural{LCA}{life cycle analyses}
\acro{LDP}{local data processing}
\acro{LIS}{large intelligent surface}
\acro{LoS}{line-of-sight}
\acro{MAC}{medium access control}
\acro{MEC}{multi-access edge computing}
\acro{MIMO}{multiple-input multiple-output}
\acro{ML}{machine learning}
\acro{mmWave}{millimeter-wave}
\acro{MTT}{multi-target tracking}   
\acro{NF}{network function}
\acro{NLoS}{non-line-of-sight}
\acro{NR}{new radio}
\acro{NTN}{non-terrestrial network}
\acro{OFDM}{orthogonal frequency-division multiplexing}
\acro{OTFS}{orthogonal time-frequency-space}
\acro{PRS}{positioning reference signal}
\acro{PHY}{physical layer}
\acro{QoS}{Quality of Service}
\acro{RAN}{radio access network}
\acro{RAT}{radio access technology}
\acro{RedCap}{reduced capacity}
\acro{RF}{radio frequency}
\acro{RIS}{reconfigurable intelligent surface}
\acro{RMSE}{root mean square error}
\acro{RTK}{real-time kinematic}
\acro{RTT}{round-trip-time}
\acro{RU}{radio unit}
\acro{SDG}{sustainable development goal} 
\acro{SDO}{standards organization}
\acro{SLAM}{simultaneous localization and mapping}
\acro{SNR}{signal-to-noise ratio}
\acro{SIT}{sustainability, inclusiveness, and trustworthiness}
\acro{SOTA}{state of the art}
\acro{SL}{sidelink}
\acro{SRN}{sensing receive node}
\acro{ToA}{time-of-arrival}
\acro{TDoA}{time-difference-of-arrival}
\acro{TR}{time-reversal}
\acro{TRP}{transmission and reception point}
\acro{TRL}{technology readiness level}
\acro{TXRX}[TX/RX]{transmitter/receiver}
\acro{TX}{transmitter}
\acro{RX}{receiver}
\acro{UE}{user equipment}
\acro{UN}{United Nations}
\acro{multi-RTT}{multi-cell round-trip-time}
\acro{UL}{uplink}
\acro{UL-TDOA}{uplink time-difference-of-arrival}
\acro{DL-TDOA}{downlink time-difference-of-arrival}
\acro{UMi}{3D-urban micro}
\acro{UMa}{3D-urban macro}
\acro{UWB}{ultra-wide band}
\acro{FR1}{frequency range 1}
\acro{FR2}{frequency range 2}
\acro{WLAN}{wireless local-area network}
\acro{XL-MIMO}{extremely large-scale MIMO}
\acro{FOV}{field-of-view}
\acro{RBPF}{Rao-Blackwellized particle filter}
\acro{PHD}{probability hypothesis density}
\acro{OID}{optimal importance density}
\end{acronym}